\documentclass[
    ,final            
  ]
  {aipproc}

\layoutstyle{6x9}

\begin{document}

\title{Spin Observables for Polarizing Antiprotons}



\author{D.~S.~O'Brien}{
  address={School of Mathematics, Trinity College Dublin, Ireland}
}

\author{N.~H.~Buttimore}{
  address={School of Mathematics, Trinity College Dublin, Ireland}
}


\begin{abstract}
The PAX project at GSI Darmstadt plans to polarize an antiproton beam by repeated interaction with a hydrogen target in a storage ring.  Many of the beam particles are required to remain within the ring after interaction with the target, so small scattering angles are important.  Hence we concentrate on low momentum transfer (small $t$), a region where electromagnetic effects dominate the hadronic effects.  A colliding beam of polarized electrons with energy sufficient to provide scattering of antiprotons beyond ring acceptance may polarize an antiproton beam by spin filtering.  Expressions for spin observables are provided and are used to estimate the rate of buildup of polarization of an antiproton beam.
\end{abstract}

\maketitle


\section{Introduction}

The buildup of the polarization of an antiproton beam
by spin filtering off a polarized hydrogen target
has been described by differential equations
\cite{Nikolaev:2006gw,Milstein:2005bx}
involving the spin observables for antiproton-electron
and antiproton-proton elastic scattering.
Relativistic expressions for spin observables due to single
photon exchange in elastic spin 1/2 - spin 1/2  particle collisions
have been presented recently
\cite{O'Brien:2006zt}.
There has been much recent theoretical debate as to
whether electrons in a hydrogen gas target are effective
in transferring polarization to the antiproton beam
\cite{Nikolaev:2006gw,Milstein:2005bx}.
It is claimed that electrons should scatter antiprotons
out of the beam to contribute to spin filtering.
The maximum antiproton-electron scattering angle for
various antiproton beam energies is lower than the
ring acceptance angles under consideration by PAX
\cite{Rathmann:2004pm,Barone:2005pu}
and we suggest that a beam of polarized electrons with
energy sufficient to provide scattering of antiprotons
beyond ring acceptance may polarize an antiproton beam
by spin filtering.

\section{Spin observables}

For electromagnetic interactions to first order the double spin asymmetries equal the polarization transfer observables ($\, A_{ij} = K_{ij}\, $) and all the single and triple spin asymmetries are zero ($\, A_{i} = A_{ijk} = 0\, $) where $i,j,k \in \{\mathrm{\,X,Y,Z\,}\}$.

Spin filtering requires evaluation of the angular integration of the product of the observables $A_{ii} = K_{ii}$ and $\left(1 - D_{ii}\right)$ with $d\,\sigma / d\,\Omega$.  Azimuthal averaging indicates that the observables with single $\mathrm{X}$ ( i.e. $K_\mathrm{XZ}$, $K_\mathrm{ZX}$, $D_\mathrm{XZ}$ and $D_\mathrm{ZX}$ ) do not contribute to spin filtering.  The quantities $\left(K_\mathrm{XX} + K_\mathrm{YY}\right)/2$, $\left(D_\mathrm{XX} + D_\mathrm{YY}\right)/2$, $K_\mathrm{ZZ}$, $D_\mathrm{ZZ}$ and $d\,\sigma / d\,\Omega$ which we now present, play the important role.  The spin averaged differential cross-sections have been presented in \cite{O'Brien:2006zt,O'Brien:2006tp}.  In the following $m$ is the electron mass, $M$ is the (anti)proton mass, $\mu_\mathrm{p}$ is the magnetic moment of the proton, $\alpha$ is the fine structure constant, $s$ and $t$ are Mandelstam variables and
$k = \frac{1}{2}[s - 2M^2 - 2m^2 + (M^2 - m^2)^2/s]^{1/2}\!.$

\subsection{Antiproton - proton scattering}
\noindent
To leading order in small $t$ the relevant spin observables for single photon exchange antiproton-proton scattering are as follows \cite{O'Brien:2006zt}.  The $\approx$ sign refers to the first term in the expansion in $t$.
\begin{eqnarray}
\frac{K_\mathrm{XX} + K_\mathrm{YY}}{2}\ \frac{d\,\sigma}{d\,\Omega} &\approx &\displaystyle{\frac{\alpha^2\ M^{\,2}\ \mu_\mathrm{p}^{\,2}}{s\ t}} \nonumber \\[1ex]
\frac{\left(1 - D_\mathrm{XX}\right) + \left(1 - D_\mathrm{YY}\right)}{2}\ \frac{d\,\sigma}{d\,\Omega} &\approx &\displaystyle{ \frac{-\ \alpha^2 \left(k^2 + M^{\,2}\right)}{k^2\ M^{\,2}\ s\ t}\left[M^{\,2} - 2\,k^2 \left(\mu_\mathrm{p} -1\right)\right]^{\,2}} \nonumber \\[1ex]
K_\mathrm{ZZ}\ \frac{d\,\sigma}{d\,\Omega} & \approx & \displaystyle{\frac{-\,2\ \alpha^2 \ \mu_\mathrm{p}^2}{s\ t}\ \left(2\,k^2 + M^{\,2}\right)}\\[1ex]
\left(1 - D_\mathrm{ZZ}\right)\ \frac{d\,\sigma}{d\,\Omega} & \approx & \displaystyle{ \frac{-\, 2\,\alpha^2 \left(k^2 + M^{\,2}\right)}{k^2\ M^{\,2}\ s\ t}\left[M^{\,2} - 2\,k^2 \left(\mu_\mathrm{p} -1\right)\right]^{\,2}}\,.\nonumber
\end{eqnarray}

\subsection{Antiproton - electron scattering}
\noindent
The leading $t$ terms in the relevant observables for antiproton-electron scattering are \cite{O'Brien:2006zt}: 
\begin{eqnarray}
\frac{K_\mathrm{XX} + K_\mathrm{YY}}{2}\ \frac{d\,\sigma}{d\,\Omega} &\approx &\displaystyle{ \frac{\alpha^2\ m\ M\ \mu_\mathrm{p}}{s\ t}} \nonumber \\[1ex]
\frac{\left(1 - D_\mathrm{XX}\right) + \left(1 - D_\mathrm{YY}\right)}{2}\ \frac{d\,\sigma}{d\,\Omega} &\approx &\displaystyle{\frac{-m^2\,\alpha^2\,\left(\,s - m^2 + M^{\,2}\,\right)^2}{4\ k^2\ s^2\ t}} \nonumber \\[1ex]
K_\mathrm{ZZ}\ \frac{d\,\sigma}{d\,\Omega} &\approx & \displaystyle{\frac{-\, \alpha^2\ \mu_\mathrm{p}}{s\ t}\left(\,s - m^2 - M^{\,2}\,\right)}\\[1ex]
\left(1 - D_\mathrm{ZZ}\right)\ \frac{d\,\sigma}{d\,\Omega} &\approx & \displaystyle{\frac{-M^2\,\alpha^2\,\left(\,s + m^2 - M^{\,2}\,\right)^2}{2\ k^2\ s^2\ t}}\,.\nonumber
\end{eqnarray}

\section{Antiprotons scattering off stationary electrons}

The maximum scattering angle for $10 \ \mbox{GeV}/c$ antiprotons scattering off electrons in an atomic target is $0.54\ \mbox{mrad}$ as shown in Figure 1 (a).  This is below the acceptance angles under consideration at PAX \cite{Rathmann:2004pm,Barone:2005pu} so that all scattering off atomic electrons will be within the ring; therefore no spin filtering is possible and electrons will not contribute to the polarization buildup of the antiproton beam.  An opposing electron beam of sufficient energy would increase the scattering angles of the antiprotons beyond acceptance as seen in Figure 1 (b), hence allowing spin filtering.
\begin{figure}
\hspace*{0.2em}
     \includegraphics{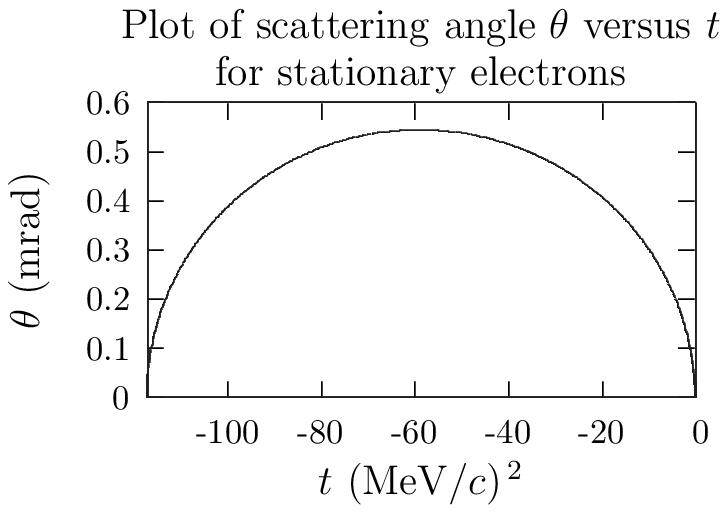}
\hspace*{-1.3em}
     \includegraphics{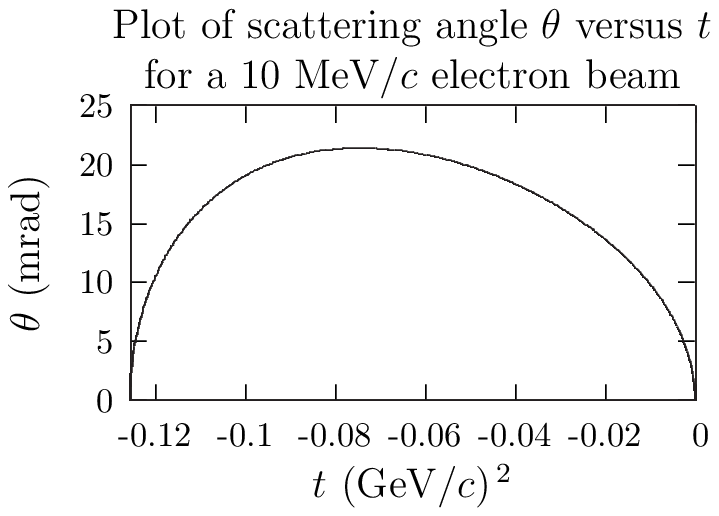}
\caption{(a) Scattering angle versus $t$ for $10 \ \mbox{GeV}/c$ antiprotons scattering off electrons in an atomic target.  (b) With a colliding electron beam the maximum scattering angle increases as the electron beam momentum increases in a direction opposite to that of the antiproton beam.
}
\end{figure}
\section{Polarization buildup}

When circulating at frequency $\nu$ through a polarized target of areal density $n$ and polarization $P_\mathrm{e}$ oriented normal to the ring plane, (or longitudinally with rotators)
\begin{eqnarray}
  \frac{d}{d\tau}
\left[
        \begin{array}{c} N \\[2ex] J \end {array}
\right]
\,  = \, - \, n \, \nu
\left[
\begin{array}{ccc}
         I_\mathrm{\, out} && P_\mathrm{e} \, A_\mathrm{\, out}
\\[2ex]
    P_\mathrm{e} \, A_\mathrm{\, all} -  P_\mathrm{e} \, K_\mathrm{in}
&&
         I_\mathrm{\, all} -  D_\mathrm{\, in}
\end {array}
\right]
\,
\left[
        \begin{array}{c} N \\[2ex] J \end {array}
\right]
\end{eqnarray}
 describes the rate of change of the number of beam particles $N$
 and their total spin $J$ \cite{Nikolaev:2006gw,Milstein:2005bx}.
 These coupled differential equations involve angular integration of
 the spin observables presented earlier, as seen in the following table.
 The minimum angle $\theta_0$ relates to the average
 transverse electron separation,
 and $\theta_\mathrm{acc}$ is the acceptance angle.
 The time ($\tau$) dependence of the polarization of the beam
 is given by solving the system 
\begin{equation}
   P(\tau) \, = \frac{J(\tau)}{N(\tau)}\,=-\, P_\mathrm{e}
\,\,
\frac{ A_\mathrm{all} \, - \, K_\mathrm{in}
}
{   L_\mathrm{in} \, + \, L_\mathrm{d} \, \coth\left(L_\mathrm{d}\, n\,\nu\,\tau\right)
}
\end{equation}
 where the discriminant of the quadratic equation
 for the eigenvalues is
\begin{equation}
   L_\mathrm{d}
\, =
\, \sqrt{\, P_\mathrm{e}^{\,2} \, A_\mathrm{out} \left( A_\mathrm{all}
\, -
\, K_\mathrm{in} \right) \, + \, L_\mathrm{in}^{\,2} }
\end{equation}
and where
$L_\mathrm{in} \, = \left( \, I_\mathrm{in} \, - \, D_\mathrm{in}\right)\,/\,2$
is a loss of polarization quantity.
The approximate rate of change of
 polarization for sufficiently short times, and the limit of the polarization for large times are respectively:
\begin{equation}
   \frac{dP}{d\tau} \, \, \approx \, -\, n \, \nu \, P_\mathrm{e}
\,
\left( A_\mathrm{all} \, - \, K_\mathrm{in} \right)
\hspace*{1.5cm}
\displaystyle{
   \lim_{\tau \to \infty} P(\tau) \, = -\, P_\mathrm{e}
\,
\frac{ A_\mathrm{all} \, - \, K_\mathrm{in}
}
{   L_\mathrm{in} \, + \, L_\mathrm{d}
}}
\,.
\end{equation}

\begin{table}
\begin{tabular}{|c|c|} 
\hline
 & \\
Transverse polarization requires & Longitudinal polarization requires \\[2ex] \hline
 & \\
$\displaystyle{
\,
   I_\mathrm{out}
\, =
\, 2 \, \pi \int_{\theta_\mathrm{acc}}^{\pi}
\,
   \frac{d\,\sigma}{d\,\Omega} \,\, \sin\theta \, d\theta
}
$
& 
$\displaystyle{
\,
   I_\mathrm{out}
\, =
\, 2 \, \pi \int_{\theta_\mathrm{acc}}^{\pi}
\,
   \frac{d\,\sigma}{d\,\Omega} \,\, \sin\theta \, d\theta
}
$\\[3ex]
$\displaystyle{
\!
    A_\mathrm{out}
\!
   =
\!
 \pi \!\!\! \int_{\theta_\mathrm{acc}}^{\pi}
\!\!\!\!\!
   \left( A_\mathrm{XX} \! + \! A_\mathrm{YY} \right)
   \frac{d\,\sigma}{d\,\Omega} \, \sin\theta \, d\theta \!
}
$
&
$\displaystyle{
\,
    A_\mathrm{out}
\,
   =
\, 2\,\pi \int_{\theta_\mathrm{acc}}^{\pi}
\!
   A_\mathrm{ZZ}\,\,
   \frac{d\,\sigma}{d\,\Omega} \,\, \sin\theta \, d\theta 
}
$\\[3ex]
$ \displaystyle{
\!
   A_\mathrm{all}
\! =
\! \pi \!\!\int_{\theta_0}^{\pi}
 \!\!  \left( A_\mathrm{XX} \! + \! A_\mathrm{YY} \right)
   \frac{d\,\sigma}{d\,\Omega} \, \sin\theta \, d\theta\!
}
$
&
$ \displaystyle{ 
\,  
   A_\mathrm{all}
\, =
\, 2\,\pi \int_{\theta_0}^{\pi}
    A_\mathrm{ZZ}\,\,
   \frac{d\,\sigma}{d\,\Omega} \,\, \sin\theta \, d\theta
}
$\\[3ex]
$\displaystyle{
\!
    K_\mathrm{in}
\!
   =
\!  \pi \!\! \!\int_{\theta_0}^{\theta_\mathrm{acc}}
\!\!\!\!\!
   \left( K_\mathrm{XX} \! + \! K_\mathrm{YY} \right)
   \frac{d\,\sigma}{d\,\Omega} \, \sin\theta \, d\theta\!
}
$
&
$\displaystyle{
\,
    K_\mathrm{in}
\,
   =
\,  2\,\pi \int_{\theta_0}^{\theta_\mathrm{acc}}
\!
  K_\mathrm{ZZ}\,\,
   \frac{d\,\sigma}{d\,\Omega} \,\, \sin\theta \, d\theta
}
$\\[3ex]
$\displaystyle{
\!
    D_\mathrm{in}
\!
   =
\!  \pi\!\!\! \int_{\theta_0}^{\theta_\mathrm{acc}}
\!
  \!\!\! \left( D_\mathrm{XX} \! + \! D_\mathrm{YY} \right)
   \frac{d\,\sigma}{d\,\Omega} \, \sin\theta \, d\theta\!
}
$
&
$\displaystyle{
\,
    D_\mathrm{in}
\,
   =
\,  2\,\pi \int_{\theta_0}^{\theta_\mathrm{acc}}
\!
  D_\mathrm{ZZ}\,\,
   \frac{d\,\sigma}{d\,\Omega} \,\, \sin\theta \, d\theta
}$
\\[3ex]\hline
\end{tabular} 
\end{table}
\pagebreak
\section{Conclusions}

We have shown that, in principle, a sufficiently high density polarized 
electron beam could be used to polarize an antiproton beam by spin filtering in a storage ring.  This can be done by a pure electromagnetic process.  The polarization buildup rate is proportional to the density of the electrons in the beam.  We are now investigating if the rate of polarization buildup and maximum polarization achievable with current technologies are sufficient to provide a useful figure of merit when attempting to polarize antiprotons.
%

\begin{theacknowledgments}

DOB would like to thank \lq\lq The Embark Initiative'' (IRCSET), for a postgraduate research scholarship.  NHB is grateful to Enterprise Ireland for the award of a grant under the International Collaboration Program to facilitate a visit to INFN at the University of Torino, where discussions with M.~Anselmino and M.~Boglione are acknowledged.  We are also grateful to E.~Leader, N.~N.~Nikolaev and F.~Rathmann for helpful comments.
\end{theacknowledgments}







\begin{thebibliography}{9}

\bibitem{Nikolaev:2006gw}
  N.~N.~Nikolaev and F.~F.~Pavlov,
  arXiv:hep-ph/0601184.

\bibitem{Milstein:2005bx}
  A.~I.~Milstein and V.~M.~Strakhovenko,
  \emph{Phys.\ Rev.\ E} {\bf 72} 066503 (2005).

\bibitem{O'Brien:2006zt}
  D.~S.~O'Brien and N.~H.~Buttimore,
  arXiv:hep-ph/0609233.

\bibitem{Rathmann:2004pm}
  F.~Rathmann \emph{et al.},
  \emph{Phys.\ Rev.\ Lett.}\  {\bf 94} 014801 (2005).

\bibitem{Barone:2005pu}
  V.~Barone \emph{et al.} [PAX Collaboration],
  arXiv:hep-ex/0505054.

\bibitem{O'Brien:2006tp}
  D.~S.~O'Brien and N.~H.~Buttimore,
  \emph{Czech.~J.~Phys.} {\bf 56} C265 (2006)
  [arXiv:hep-ph/0605099].



\end{thebibliography}
\end{document}